\documentclass[epjST]{svjour}
\usepackage{graphics}
\usepackage{graphicx}
\usepackage{amsmath}
\usepackage{amsfonts}
\usepackage{amssymb}
\begin{document}
\title{Computational efficiency of symplectic integration schemes:
  Application to multidimensional disordered Klein-Gordon lattices}
\subtitle{}
\author{B. Senyange\inst{1}\fnmsep\thanks{\email{snybob001@myuct.ac.za}}
  \and Ch. Skokos\inst{1}\fnmsep\thanks{\email{haris.skokos@uct.ac.za}}}
\institute{Department of Mathematics and Applied Mathematics,
  University of Cape Town, \\ Rondebosch, 7701, Cape Town, South
  Africa.}
\abstract{We implement several symplectic integrators, which are based
  on two part splitting, for studying the chaotic behavior of one- and
  two-dimensional disordered Klein-Gordon lattices with many degrees
  of freedom and investigate their numerical performance.  For this
  purpose, we perform extensive numerical simulations by considering
  many different initial energy excitations and following the
  evolution of the created wave packets in the various dynamical
  regimes exhibited by these models.  We compare the efficiency of the
  considered integrators by checking their ability to correctly
  reproduce several features of the wave packets propagation, like the
  characteristics of the created energy distribution and the time
  evolution of the systems' maximum Lyapunov exponent estimator.
  Among the tested integrators the fourth order $ABA864$ scheme
  \cite{BCFLMM13} showed the best performance as it needed the least
  CPU time for capturing the correct dynamical behavior of all
  considered cases when a moderate accuracy in conserving the systems'
  total energy value was required. Among the higher order schemes used
  to achieve a better accuracy in the energy conservation, the sixth
  order scheme $s11ABA82\_6$ exhibited the best performance.}
%
\maketitle
%
\section{Introduction}
\label{sec:intro}

Disordered dynamical systems try to model heterogeneity appearing in
nature due to e.g.~impurities, imperfections and defects. Typically,
disorder is introduced by giving random, uncorrelated values to one or
more parameters of a system. A fundamental phenomenon in disordered
systems, which is usually referred as `Anderson localization', is the
halting of energy propagation in the presence of sufficiently strong
disorder \cite{A58}.  The appearance and/or the destruction of
Anderson localization in linear and nonlinear disordered systems, as
well as the properties of energy propagation in such models, have
attracted extensive attention in theory, numerical simulations and
experiments, especially in recent years
\cite{S93,M98,CVHRBSGSA05,FFGLMWI05,SBFS07,LAPSMCS08,BJZBHLCS08,REFFFZMMI08,KKFA08,PS08,FKS09,FKS09b,GMS09,SKKF09,VKF09,SF10,LBKSF10,F10,B11,MAP11,MAPS11,BLSKF11,BLGKSF11,KMZD11,LBF12,MP12,JBMCJPPSAB12,MP13,SGF13,ABSD14,LIF14,SKKF14,TSL14,MKP16,ATS16,MYKKPY6,ASBF17,CPKD17,DHKBB17,ATS17}.

Studies of disordered versions of two basic, nonlinear Hamiltonian
lattice models, namely the Klein-Gordon (KG) oscillator lattice and
the discrete nonlinear Schr\"{o}dinger equation (DNLS), revealed the
existence of various dynamical behaviors, the so-called `weak' and
`strong chaos' spreading regimes, as well as the `selftrapping' regime
and determined the statistical characteristics of energy propagation
and chaos in these systems
\cite{FKS09,FKS09b,SKKF09,SKKF14,SF10,LBKSF10,F10,BLSKF11,BLGKSF11,LBF12,SGF13,LIF14}. One
basic outcome of these studies is that energy propagation in
disordered lattices is a chaotic process which, in general, results in
the destruction of Anderson localization.

Although our understanding of the mechanisms which govern energy
spreading in disordered nonlinear lattices has improved significantly
over the last decade, many important questions concerning mainly the
asymptotic behavior of wave packets and the effect of chaos on that
behavior are still open. The numerical investigation of these
questions requires the accurate integration of Hamiltonian models with
many degrees of freedom (of the order of a few thousands) for very
long times and for several different disorder realizations in order to
obtain solid and reliable statistical results. This is a very
demanding computational task, which will be greatly benefited from the
use of efficient integration techniques which could allow the accurate
long time integration of multidimensional Hamiltonians in feasible CPU
times.

Symplectic integrators (SIs) are numerical techniques explicitly
designed for the integration of Hamiltonian systems and nowadays are
used widely for this purpose (see for example
\cite{HLW02,MQ02,MQ06,F06,BCM08} and references therein). A basic
characteristic of these integrators is that they keep the error of the
value of the Hamiltonian, i.e.~the usually called `energy', bounded as
time increases, in contrast to non-symplectic integrators for which
the error increases in time. This feature is of particular interest
for the long-time integration of disordered nonlinear lattices, as it
guarantees the accurate computation of the asymptotic behavior of such
systems. In addition, the fact that SIs can achieve this accuracy for
relatively large integration time steps results to the decrease of the
required CPU time.

For all these reasons SIs based on splitting the system's Hamiltonian
in distinct, integrable parts have already been used for the
integration of the KG and DNLS models
\cite{FKS09,SKKF09,SF10,LBKSF10,BLSKF11,LBF12,SGF13,ABSD14,TSL14,SGBPE14,GMS16,ASBF17}. In
these studies SIs belonging mainly in the so-called SABA schemes
\cite{LR01} were used. The numerical integration of the KG system,
both in one and two dimensions, proved to be computationally easier as
the KG models can be split in two integrable parts (the kinetic and
the potential energy), while the efficient integration of the DNLS
models requires the splitting of the corresponding Hamiltonian in
three integrable parts \cite{BLSKF11,SGBPE14,GMS16}. As a result,
using the same computational resources and CPU times one can integrate
the KG models for times of one or two orders of magnitude higher than
the times achieved for the DNLS systems. For this reason ways to
improve the efficiency of the symplectic integration of the DNLS model
were investigated in \cite{SGBPE14,GMS16} where the performance of
several SIs based on three part splits, with different orders of
accuracy, were studied in detail.

A similar analysis for the KG model is lacking, and it is exactly this
gap that the current paper fills. In our study we consider a plethora
of SIs of various orders not only for the integration of the Hamilton
equations of motion, which govern the evolution of orbits in the
system's phase space, but also for the simultaneous integration of the
so-called `variational equations', governing the time evolution of
small deviation vectors from the studied orbit. These deviation
vectors are needed for characterizing the system's chaoticity through
the computation of a chaos indicator \cite{SGL16}, like the maximum
Lyapunov Characteristic Exponent (mLCE) \cite{BGGS80a,BGGS80b,S10} we
consider in this work. We note that our investigation includes both
the one- and two-dimensional KG models.

The paper is organized as follows: In Sect.~\ref{sec:kg} we describe
the two Hamiltonian models we consider in our study, while in
Sect.~\ref{sec:si} after a brief introduction of the basic properties
of SIs we provide detailed information for the symplectic schemes we
include in our investigation. Then, in Sect.~\ref{sec:numres} we
present our numerical results on the performance of the various SIs we
implemented. Finally, in Sect.~\ref{sec:conclusions} we summarize our
findings. In the Appendix the explicit form of several operators needed
for the symplectic integration of both
the one- and two-dimensional KG models are provided.

\section{The Klein-Gordon lattice models}
\label{sec:kg}

The one-dimensional (1D) KG lattice model of $N$ coupled anharmonic
oscillators is described by the Hamiltonian
\begin{equation}
\label{eq:ham1}
 H_1(\vec{q},\vec{p}) = \sum_{i=1}^N \left[ \frac{p_i^2}{2} +
   \frac{\epsilon_i}{2}q_i^2 + \frac{q_i^4 }{4}+
   \frac{1}{2W}\left(q_{i+1} - q_i\right)^2 \right],
\end{equation}
where $\vec{q}= (q_1, q_2, \ldots, q_N)$ and $\vec{p}= (p_1, p_2,
\ldots, p_N)$ are respectively the generalized positions (representing
displacements of oscillators from their equilibriums) and momenta,
$\epsilon_i$ are parameters chosen uniformly from the interval
$[\frac{1}{2},\frac{3}{2}]$, which determine the on-site potentials,
while $W$ denotes the disorder strength. We also consider fixed
boundary conditions for this lattice, i.e.~$q_{N+1}=0$.

A natural, simple, extension of this model in two dimensions is
obtained by attributing a scalar displacement, $q_{i,j}$, to each
lattice site of a two-dimensional orthogonal arrangement of $N\times
M$ oscillators having $N$ nodes in one direction (related to index
$i$) and $M$ nodes in the other (related to index $j$). Then, the
corresponding 2D KG Hamiltonian is
\begin{equation}
\label{eq:ham2}
H_2(\vec{q},\vec{p}) = \sum_{j=1}^M\sum_{i=1}^N \left\{
\frac{p^2_{i,j}}{2} + \frac{\epsilon_{i,j}}{2} q^2_{i,j} +
\frac{q_{i,j} ^{4}}{4} + \frac{1}{2W}\left[(q_{i+1,j} - q_{i,j})^2 +
  (q_{i,j+1} - q_{i,j})^2\right] \right\},
\end{equation}
where $p_{i,j}$ is the generalized conjugate momentum of oscillator
$(i,j)$ and $\epsilon_{i,j}$ are again chosen uniformly from the
interval $[\frac{1}{2},\frac{3}{2}]$. Again fixed boundary conditions
are imposed so that $q_{N+1,j}=q_{i,M+1}=0$ for $i=1,2,\ldots,N$,
$j=1,2, \ldots, M$.

The dynamics of Hamiltonian
(\ref{eq:ham1}) was studied in
\cite{FKS09,SKKF09,LBKSF10,BLSKF11,SGF13,ABSD14,ASBF17}, while system
(\ref{eq:ham2}) was considered in \cite{LBF12}.

\section{Symplectic integration schemes}
\label{sec:si}

The equations of motion of a Hamiltonian system $H(\vec{q},\vec{p})$
with $m$ degrees of freedom can be written as ${\frac{d \vec{z}}{dt}=
  \{\vec{z},H\} = L_H\vec{z}}$, where $\vec{z} = (\vec{q},\vec{p})$
and $L_H=\{\cdot, H\}$ is a differential operator with $\{\cdot,
\cdot \}$ being the Poisson bracket defined by $\{
F,G\}=\sum_{i=1}^{m} \left( \frac{\partial F}{\partial q_i}
\frac{\partial G}{\partial p_i} - \frac{\partial F}{\partial p_i}
\frac{\partial G}{\partial q_i}\right)$ for any differentiable
functions $F(\vec{z})$ and $G(\vec{z})$. The vector $\vec{z}$
corresponds to a point in the $2m$-dimensional phase space of the
system, while its time evolution $\vec{z}(t)$ determines an orbit in
that space.  Using initial conditions $\vec{z}(t_0)$ at time $t=t_0$
we can formally write the solution $\vec{z}(t_0+\tau)$ of the Hamilton
equations of motion at time $t=t_0+\tau$ as
$\vec{z}(t_0+\tau)=\sum_{i\geq 0} \frac{\tau^i}{i!}  L_H^i
\vec{z}(t_0)=e^{\tau L_H} \vec{z}(t_0)$. So $e^{\tau L_H}$ is the
operator which propagates in time the coordinate vector $\vec{z}$ by
$\tau$ time units. In general, the action of this operator is not known
analytically.

In many cases the Hamiltonian function $H(\vec{z})$ can be written as
a sum of two integrable parts, $H(\vec{z})=A(\vec{z})+B(\vec{z})$, so
that the action of operators $e^{\tau L_A}$ and $e^{\tau L_B}$ is
known analytically. An explicit SI of order $n$, $n \in
\mathbb{N}$,  approximates the
action of operator $e^{\tau L_H}$ by a series of products of operators
$e^{a_i \tau L_A}$ and $e^{b_i \tau L_B}$, i.e.
\begin{equation}\label{eq:prod}
    e^{\tau L_H}=e^{ \tau \left( L_A+L_B \right)}=\prod_{i=1}^p e^{a_i
      \tau L_A} e^{b_i \tau L_B} + \mathcal{O}(\tau^{n+1}),
\end{equation}
where $a_i$, $b_i$, $i=1,2,\ldots,p$, are appropriately chosen
constants for obtaining the desired order of accuracy. Usually the
total number of applications of the simple operators $e^{a_i \tau
  L_A}$ and $e^{b_i \tau L_B}$ is referred as the number of `steps' of
the integrator.

Both the 1D (\ref{eq:ham1}) and the 2D (\ref{eq:ham2}) KG
Hamiltonians, $H_i(\vec{q},\vec{p})$, $i=1,2$, can be written as a sum
of two integrable parts: the kinetic energy $A_i(\vec{p})$, which
depends only on the system's momenta, and the potential energy
$B_i(\vec{q})$, which is a function of only the generalized positions,
i.e.~$H_i(\vec{q},\vec{p}) = A_i(\vec{p})+ B_i(\vec{q})$.  Thus, SIs
based on two part splitting can be applied straightforwardly for the
numerical integration of systems (\ref{eq:ham1}) and (\ref{eq:ham2}).

Over the years several SIs of various orders and different number of
steps have been developed and implemented. SIs of higher order achieve
better accuracy for the same integration time step $\tau$, but their
implementation could require more computational effort as they contain
more steps than low order schemes. In our study we consider in total
33 SIs whose order ranges from 2 up to 8. In the remaining part of
this section we briefly present all these SIs.

\subsection{Symplectic integrators of order two}
\label{sec:SI2}

The simplest symmetric SI of order two is the so-called `leap frog'
integrator ($LF$) or Verlet integrator (see for example \cite{R83} and
\cite[Sect.~I.3.1]{HLW02}), having 3 steps
\begin{equation}
\label{eq:LF}
 LF(\tau) = e^{a_1\tau L_A}e^{b_1\tau L_B}e^{a_2\tau L_A},
\end{equation}
with $a_1 = a_2 = \frac{1}{2}$ and $b_1 = 1$. In our study we also
consider the second order, 5 step schemes $SABA_2$, $SBAB_2$
\cite{LR01} having respectively the forms
\begin{equation}
\label{eq:SABA2}
SABA_2(\tau)=e^{a_1\tau L_A}e^{b_1\tau L_B}e^{a_2\tau L_A}e^{b_1\tau L_B}e^{a_1\tau L_A},
\end{equation}
with $a_1=\frac{1}{2} - \frac{1}{2\sqrt{3}}$, $a_2 =
\frac{1}{\sqrt{3}}$, $b_1 = \frac{1}{2}$, and
\begin{equation}
\label{eq:SBAB2}
SBAB_2(\tau) = e^{b_1\tau L_B}e^{a_1\tau L_A}e^{b_2\tau L_B}e^{a_1\tau L_A}e^{b_1\tau L_B},
\end{equation}
for $a_1 = \frac{1}{2}$, $b_1 = \frac{1}{6}$ and $b_2 = \frac{2}{3}$,
as well as the $ABA82$ SI \cite{M95,FLBCMM13} of 9 steps
\begin{equation}
\label{eq:ABA82}
ABA82(\tau) = e^{a_1\tau L_A}e^{b_1\tau L_B}e^{a_2\tau L_A}e^{b_2\tau
  L_B}e^{a_3\tau L_A}e^{b_2\tau L_B} e^{a_2\tau L_A}e^{b_1\tau
  L_B}e^{a_1\tau L_A},
\end{equation}
whose coefficients $a_i$, $b_i$, $i=1,2,3$ can be found
in Table 2 of \cite{FLBCMM13}. We note that the $ABA82$ integrator is
called $SABA_4$ in \cite{LR01}.

\subsection{Symplectic integrators of order four}
\label{sec:SI4}

A way to construct higher order SIs is through a composition
(i.e.~successive application) of lower order schemes. A composition
approach of this kind was proposed in \cite{Y90}. According to that
method, starting from a SI $S_{2n}(\tau)$ of order $2n$ we construct a SI $S_{2n+2}(\tau)$ of order $2n+2$ as
\begin{equation}
\label{eq:CompYosh}
    S_{2n+2}(\tau)=S_{2n}(d_1\tau)S_{2n}(d_0\tau)S_{2n}(d_1\tau),
\end{equation}
where $d_0=-2^{1/(2n+1)} / [2-2^{1/(2n+1)}]$ and $d_1=1
/[2-2^{1/(2n+1)}]$. We note that if we start from a second order
integrator $S_2$ then the construction of the SI $S_{2n+2}$ requires
the application of $S_2$ $3^{n}$ times. Obviously with this approach
the number of steps of the $S_{2n+2}$ scheme grows rapidly when $n$
increases, despite the fact that adjacent applications of the same
basic operators $e^{a_i\tau L_A}$, or $e^{b_i\tau L_B}$ can be grouped
together.

Starting from a SI of order two we can apply the composition technique
(\ref{eq:CompYosh}) for $n=2$ and create a SI of order four. If the
second order SI used is the $LF$ scheme (\ref{eq:LF}) the created SI
has 7 steps and was introduced in \cite{FR90,Y90}. We denote this
integrator by $FR4$. Its form is
\begin{equation}
    FR4(\tau)=
    e^{a_1 \tau L_A}
    e^{b_1 \tau L_B}
    e^{a_2 \tau L_A}
    e^{b_2 \tau L_B}
    e^{a_2 \tau L_A}
    e^{b_1 \tau L_B}
    e^{a_1 \tau L_A},
\label{eq:FR}
\end{equation}
with
$a_1=\frac{1}{2\left(2-2^{1/3}\right)}$,
$a_2=\frac{1-2^{1/3}}{2\left(2-2^{1/3}\right)}$, $b_1=
\frac{1}{2-2^{1/3}}$, $b_2= -\frac{2^{1/3}}{2-2^{1/3}}$.  Applying the
composition (\ref{eq:CompYosh}) to the second order integrators
$SABA_2$ (\ref{eq:SABA2}), $SBAB_2$ (\ref{eq:SBAB2}) and $ABA82$
(\ref{eq:ABA82}) we obtain the fourth order schemes $SABA_2Y4$,
$SBAB_2Y4$ and $ABA82Y4$, having 13, 13 and 25 steps respectively.

In \cite{LR01} it was shown that if the double Poisson bracket $\{
B,\{ B,A \}\}$ leads to an expression which can be seen as an
integrable Hamiltonian, then the accuracy of the $SABA_2$
(\ref{eq:SABA2}) and the $SBAB_2$ (\ref{eq:SBAB2}) integrators can be
improved by applying a corrector term $C(\tau)=e^{-\tau^3 \frac{c}{2}
  L_{\{ B,\{ B,A \}\}}}$ before and after the application of the main
body of these integrators. We note that $c=\frac{(2-\sqrt{3})}{24}$
for $SABA_2$ and $c=\frac{1}{72}$ for $SBAB_2$. For the KG
Hamiltonians (\ref{eq:ham1}) and (\ref{eq:ham2}) $\{ B,\{ B,A \}\}$
depends only on the generalized positions, so it corresponds to an
integrable system and the corrector term $C(\tau)$ can be written
analytically. Thus, applying this corrector we get two SIs of order
four, which we name $SABA_2C$ and the $SBAB_2C$, having 7 steps each.

In addition, we consider in our analysis the fourth order SIs $ABA864$
and $ABAH864$ introduced in \cite{FLBCMM13,BCFLMM13}, which have 15
and 17 steps respectively. The values of the coefficients $a_i$, $b_i$ for the $ABA864$ and $ABAH864$ integrators can be found in Tables 3 and 4
of \cite{BCFLMM13} respectively.

\subsection{Symplectic integrators of order six}
\label{sec:SI6}

Applying the composition technique (\ref{eq:CompYosh}) with $n=4$ to
the fourth order SIs $FR4$, $SABA_2Y4$, $SBAB_2Y4$, $ABA82Y4$,
$SABA_2C$ and $ABA864$ we construct the sixth order integrators
$FR4Y6$, $SABA_2Y4Y6$, $SBAB_2Y4Y6$, $ABA82Y4Y6$, $SABA_2CY6$ and
$ABA864Y6$ having 19, 37, 37, 73, 19 and 43 steps respectively.

In \cite{Y90} a composition technique for creating a sixth order SI
$S_6(\tau)$, starting from a second order SI $S_2(\tau)$ was
presented. This composition has the form
\begin{equation}
\label{eq:CY6}
 S_6(\tau)=S_2(w_3\tau)S_2(w_2\tau)S_2(w_1\tau)S_2(w_0\tau)
 S_2(w_1\tau)S_2(w_2\tau)S_2(w_3\tau),
\end{equation}
and requires less steps than the successive application of
(\ref{eq:CompYosh}) first with $n=2$ and then with $n=4$. In
\cite{Y90} three different sets of coefficients $w_i$, $i=0,1,2,3$
were given. In our study we implement the set described as `solution
A' in Table 1 of \cite{Y90} because according to \cite{M95} it shows
the best performance among the composition schemes presented in
\cite{Y90}.  This set corresponds to the composition method named
$s7odr6$ in \cite{KL97}.  Applying the composition scheme
(\ref{eq:CY6}) to the second order integrators $SABA_2$
(\ref{eq:SABA2}), $SBAB_2$ (\ref{eq:SBAB2}) and $ABA82$
(\ref{eq:ABA82}) we obtain the sixth order schemes $SABA_2Y6$,
$SBAB_2Y6$ and $ABA82Y6$, having 29, 29 and 57 steps respectively.

We also consider the composition scheme named $s9odr6b$ in
\cite{KL97}, which is based on 9 successive applications of
$S_2(\tau)$ in order to create a SI  of order
six
\begin{equation}
\label{eq:C6KL}
s9odr6b(\tau)=S_2(\delta_1\tau)S_2(\delta_2\tau)S_2(\delta_3\tau)S_2(\delta_4\tau)
S_2(\delta_5\tau) S_2(\delta_4\tau) S_2(\delta_3\tau)
S_2(\delta_2\tau)S_2(\delta_1\tau),
\end{equation}
as well as, an 11 stage composition method presented in \cite{SS05}, which we call $s11odr6$
\begin{equation}
\label{eq:C6SS}
s11odr6(\tau)=S_2(\gamma_1\tau)S_2(\gamma_2\tau)\cdots
S_2(\gamma_5\tau)S_2(\gamma_6\tau)S_2(\gamma_5\tau) \cdots
S_2(\gamma_2\tau)S_2(\gamma_1\tau).
\end{equation}
The values of
$\delta_i$, $i=1,\ldots,5$ of (\ref{eq:C6KL}) can be found in the
appendix of \cite{KL97}, while the values $\gamma_i$, $i=1,\ldots,6$
of (\ref{eq:C6SS}) are reported in Section 4.2 of \cite{SS05}. Using
$SABA_2$ (\ref{eq:SABA2}) as the second order SI $S_2$ in
(\ref{eq:C6KL}) and (\ref{eq:C6SS}) we get two SIs of order six, which
we call $s9SABA_26$ and $s11SABA_26$, having 37 and 45 steps
respectively, while putting in (\ref{eq:C6KL}) and (\ref{eq:C6SS}) the
$ABA82$ (\ref{eq:ABA82}) integrator we get the sixth order SIs
$s9ABA82\_6$ and $s11ABA82\_6$, having 73 and 89 steps respectively.

\subsection{Symplectic integrators of order eight}
\label{sec:SI8}

Finally we include in our study some SIs of order eight which are
based on appropriate compositions of a second order integrator $S_2$.

First we consider the composition scheme presented in \cite{Y90},
which contains 15 applications of $S_2$. In particular, we implement
the scheme referred as `solution D' in Table 2 of \cite{Y90} because,
according to \cite{M95,SS05} it performs better than the other
composition schemes of \cite{Y90}. Using as $S_2$ in this scheme the
$SABA_2$ (\ref{eq:SABA2}) integrator we end up with an eight order SI,
which we name $SABA_2Y8$, having 61 steps, while the use of $ABA82$
(\ref{eq:ABA82}) in the place of $S_2$ results to the $ABA82Y8$ scheme
having 121 steps.

In addition, we consider the composition scheme $s15odr8$ of
\cite{KL97} having 15 applications of $S_2$ and the 19 stage method
presented in Section 4.3 of \cite{SS05}, which requires 19
applications of $S_2$. We call the latter scheme $s19odr8$. Using
$SABA_2$ (\ref{eq:SABA2}) in the place of $S_2$ for both these
techniques we get the eight order SIs $s15SABA_28$ and $s19SABA_28$,
having 61 and 77 steps respectively, while the use of the second order
SI $ABA82$ (\ref{eq:ABA82}) gives two integrators of order eight,
which we name $s15ABA82\_8$ and $s19ABA82\_8$, having 121 and 153 steps
respectively.

\section{Numerical results}
\label{sec:numres}

In order to investigate the performance of the various SIs presented
in Sect.~\ref{sec:si} we follow the time evolution of different
initial excitations of Hamiltonians (\ref{eq:ham1}) and
(\ref{eq:ham2}) by numerically solving their equations of motion. The
efficiency of the considered SIs is checked by testing their ability
to correctly reproduce several characteristics of the resulting energy
propagation. In addition, we quantify the systems' chaoticity by evaluating the
mLCE, which is the most commonly used chaos indicator. For this
purpose we use the various SIs to also integrate the systems'
variational equations, which govern, at first order approximation, the
time evolution of an infinitesimal perturbation (usually refereed as a
deviation vector) from a considered orbit in the systems' phase
space. This vector is needed for the computation of the mLCE
$\Lambda$, because $\Lambda$ can be estimated as the limit for $t
\rightarrow \infty$ of the quantity
\begin{equation}
\label{eq:Lyap}
    L(t)=\frac{1}{t} \ln \left( \frac{\| \vec{w}(t_0+t) \| }{\|
      \vec{w}(t_0) \|}\right),
\end{equation}
often called finite time mLCE, i.e.~$\Lambda= \lim_{t \rightarrow
  \infty} L(t)$ \cite{BGGS80a,BGGS80b,S10}. In (\ref{eq:Lyap})
$\vec{w}(t_0+t)$ and $\vec{w}(t_0)$ are deviation vectors from the
studied orbit at times $t_0$ and $t_0+t$ respectively ($t > 0$),
while $\| \cdot \|$ denotes the usual vector norm. For autonomous
Hamiltonian systems like (\ref{eq:ham1}) and (\ref{eq:ham2}), $L(t)$
converges to a positive value for chaotic orbits, while for regular
orbits it tends to zero as $L(t) \propto t^{-1}$ \cite{BGS76,S10}.

For a Hamiltonian system $H(\vec{z})=H(\vec{q},\vec{p})$ with $m$
degrees of freedom an initial deviation vector
$\vec{w}(t_0)=(\vec{\delta z} (t_0))=(\vec{\delta q} (t_0), \vec{\delta
  p}(t_0))$ having as coordinates small changes from an orbit's
initial conditions evolves in time according to the variational
equations
\begin{equation}
\frac{d \vec{w}(t)}{d t} =
\left[
    \textbf{J}_{2m}\cdot\textbf{D$^2_H$}(\vec{z}(t))
\right] \cdot \vec{w}(t_0),
\label{eq:var}
\end{equation}
where $\textbf{J}_{2m}= \left[ \begin{array}{cc} \textbf{0}_{m} &
    \textbf{I}_{m} \\ -\textbf{I}_{m} & \textbf{0}_{m}
\end{array}
 \right]$ with $\textbf{I}_{m}$ being the $m\times m$ identity matrix
and $\textbf{0}_{m}$ being the $m\times m$ zero matrix, and
$\textbf{D$^2_H$}(\vec{z}(t))$ is a $2m \times 2m$ matrix with
elements $\textbf{D$^2_H$}(\vec{z}(t))_{i,j} = \left. \frac{\partial^2
  H}{\partial z_i \partial z_j}\right|_{\vec{z}(t)}$,
$i,j=1,2,\ldots,2m$. Since the elements of $\textbf{D$^2_H$}(\vec{z}(t))$ explicitly depend
on the system's orbit, the variational equations (\ref{eq:var}) have
to be integrated together with the Hamilton equations of motion.
According to the so-called `tangent map method' \cite{SG10,GS11,GES12}
this task can be performed by using symplectic integration schemes
which are appropriately extended to integrate both sets of
differential equations together.  From (\ref{eq:prod}) we see that the
dynamics induced by Hamiltonian $H$ can be approximated by successive
applications of the dynamics produced by the integrable Hamiltonians
$A$ and $B$ through the application of operators $e^{a_i \tau L_A}$
and $e^{b_i \tau L_B}$. This decomposition of the dynamics can be
extended also to the evolution of deviation vectors through the
successive applications of generalized operators which propagate in
time both the orbit and the deviation vector under the action of
Hamiltonians $A$ and $B$. We denote these operators by $e^{a_i \tau
  L_{AV}}$ and $e^{b_i \tau L_{BV}}$. The explicit form of these
operators for Hamiltonians (\ref{eq:ham1}) and (\ref{eq:ham2}) is
given in Appendix \ref{sec:ap}.

\subsection{One-dimensional KG model}
\label{sec:1d}

In our numerical simulations we create a disorder realization for the
1D KG model (\ref{eq:ham1}) having $N=1000$ sites and we follow the
evolution of different initial excitations up to a final integration
time $t_f=10^7$. Previous studies
\cite{FKS09,SKKF09,LBKSF10,F10,BLSKF11} shown that the 1D KG model can
exhibit three main dynamical regimes: the weak chaos, strong chaos and
selftrapping regimes, depending on the choice of different parameter
values and initial excitations. In order to investigate the potential
influence of these regimes on the performance of the various SIs we
study six different cases of initial excitations for $t_0=0$. In
particular, we consider the following cases.\\
Case A: we perform a
single site excitation of a site at the middle of the lattice for
$W=4$ and total energy $H_{1A}=0.4$.\\
Case B: we initially excite
$N_I=37$ adjacent sites at the middle of the lattice for $W=3$ and
total energy $H_{1B}=0.37$, so that the energy per initially exited
particle is $H_{1B}/N_I=0.01$.\\
Case C: initial excitation of
$N_I=21$ central sites for $W=4$ and total energy $H_{1C}=4.2$ (energy
per initially exited particle $H_{1C}/N_I=0.2$).\\
Case D: excitation
of a single, central site for $W=4$ and total energy
$H_{1D}=1.5$. \\
Case E: initial excitation of $N_I=100$ central sites
for $W=4$ and total energy $H_{1E}=1$.\\
Case F: all sites are
initially excited for $W=4$ and total energy $H_{1F}=10$.\\
We note
that in cases of multiple site initial excitations the same amount of
energy is given to each excited site as kinetic energy. This is done
by choosing the same, appropriate value for the momentum of the
initially excited sites, having a random sign for each site, while all
other momenta and positions are set to zero.

For each case we consider normalized energy distributions
\begin{equation}\label{eq:Ei}
    E_i=\left[ \frac{p_i^2}{2} + \frac{\epsilon_i q_i^2}{2} + \frac{q_i^4}{4} +
\frac{(q_{i+1}-q_i)^2}{4W} \right]/ H_1\, , \,\,\, i=1,2,\ldots,N,
\end{equation}
and evaluate their second moment $m_2=\sum_{i=1}^N (i-\bar{i})^2E_i$
and participation number $P=1/ \sum_{i=1}^N E_i^2$, with
$\bar{i}=\sum_{i=1}^N i E_i$. The efficiency of each SI is evaluated
by checking its ability to correctly reproduce the dynamics of the
energy propagation. For this reason we look at the shape of the
computed energy profiles, as well as at the time evolution of
$m_2(t)$, $P(t)$ and $L(t)$. For the computation of the mLCE we
consider in each studied case a random initial deviation vector having
nonzero coordinates only for the initially excited sites. In addition
we quantify the accuracy of our computations by registering the time
evolution of the absolute relative energy error $E_r(t)=\left| \right[
  H_1(t) -H_1(t_0=0)\left]/H_1(t_0=0) \right|$.

Cases A and B belong to the weak chaos regime, case C to the strong
chaos regime, while case D is a representative case of the
selftrapping behavior. For all these cases the energy does not reach
the lattice's boundaries up to the final integration time $t_f$ of our
simulations, because we want to mimic energy propagation in an
infinite lattice. Cases E and F correspond to extended initial
excitations, which can reach the fixed boundaries of the lattice
during our simulations. We considered these two cases in order to test
the performance of the integration schemes also for some general
excitations where the majority or even all sites eventually become
excited.

The 33 considered SIs in our study have different orders and various
numbers of steps.  In order to compare their efficiency in correctly
capturing the dynamics of system (\ref{eq:ham1}) we adjust the
integration time step $\tau$ of each scheme to achieve practically the
same level of accuracy. A typically acceptable level of accuracy in
numerical investigations of multidimensional disordered systems
correspond to values $E_r \lesssim 10^{-4}$
\cite{FKS09,SKKF09,LBKSF10,F10,BLSKF11}. Trying to improve a bit this
accuracy we report in Table \ref{tab:1} the values of $\tau$ which
set the obtained energy accuracy at $E_r \approx 10^{-5}$.
\begin{table}
\caption{Information on the performance of several SIs of order $n$
  used for the integration of the equations of motion and the
  variational equations of the 1D KG model (\ref{eq:ham1}) up to final
  time $t_f=10^7$ for the initial excitation of case B (see text for
  more details). The number of steps of each SI is given along with
  the integration time step $\tau$ used for obtaining an absolute
  relative energy error $E_r \approx 10^{-5}$. The required CPU time
  in seconds $T_C$ needed for each integrator is also reported. All
  simulations were performed on an Intel Xeon E5-2623 with 3.00 GHz.}
\label{tab:1}
\centerline{
\begin{tabular}{llrrr|llrrr}
\hline\noalign{\smallskip}
SI&$n$&Steps& $\tau$ & $T_C$  & SI& $n$ &Steps & $\tau$  & $T_C$  \\
\noalign{\smallskip}\hline\noalign{\smallskip}
$ABA82$&2 &9& 0.04 & 8528& $SABA_2Y6$&6 &29& 0.55 & 1402\\
$SABA_2$& 2&5& 0.02 & 12779& $s9SABA_26$&6 &37& 0.67 & 1406\\
$SBAB_2$&2 &5& 0.02 & 14431 & $ABA864Y6$& 6&43& 0.65 & 1652\\
$LF$& 2&3& 0.01 & 32280 &  $SBAB_2Y6$& 6&29& 0.46 & 1747 \\
$ABA864$&4 &15& 0.56 & 840&$s9ABA82\_6$&6 &73& 0.93 & 1920\\
$ABAH864$&4 &17& 0.38 & 1349&$FR4Y6$&6 &19& 0.18 & 3090\\
$ABA82Y4$&4 &25& 0.26 & 2629& $SABA_2CY6$&6 &19& 0.37 & 3238\\
$SABA_2C$&4 &7& 0.19 & 3351&  $SABA_2Y4Y6$&6 &37& 0.28 & 3366\\
$SABA_2Y4$&4 &13& 0.12 & 3560 &$SBAB_2Y4Y6$&6 &37& 0.28 & 3846\\
$FR4$&4 &7& 0.09 & 3310 & $SABA_2Y8$&8 &61& 0.20 & 7294\\
$SBAB_2Y4$&4 &13& 0.12 & 3835 & $ABA82Y8$&8 &121& 0.22 & 12474\\
$SBAB_2C$&4 &7& 0.14 & 4778& && &  &  \\
\noalign{\smallskip}\hline
\end{tabular}
}
\end{table}

\begin{figure}
\centerline{
\includegraphics[scale=1.0]{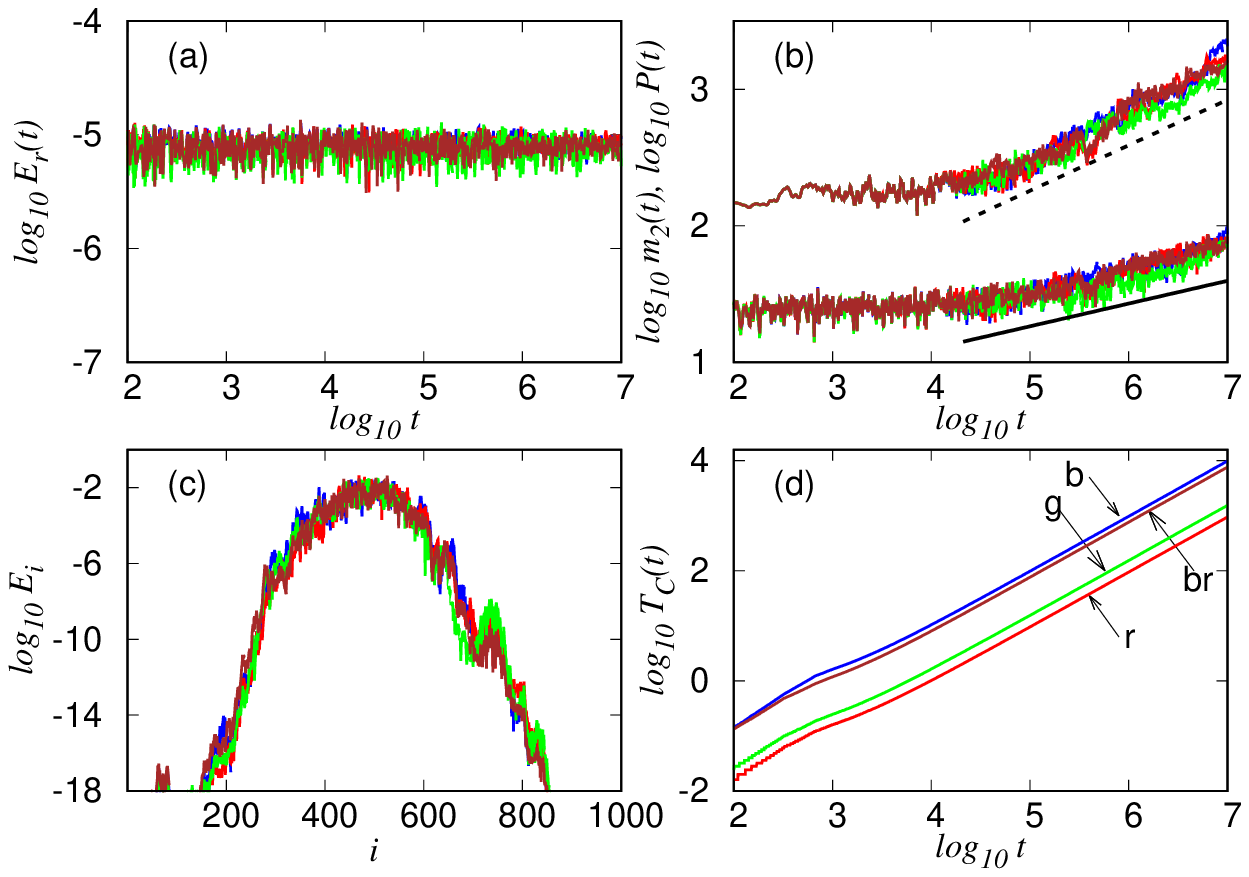}
}
\caption{Results for the integration of case B of Hamiltonian
  (\ref{eq:ham1}) by the SIs $ABA82$ of order two, $ABA864$ of order four,
  $SABA_2Y6$ of order six and $SABA_2Y8$ of order eight [(b) blue; (r) red;
    (g) green; (br) brown]: (a) the time evolution of the absolute
  relative energy error $E_r(t)$, (b) the time evolution of the second
  moment $m_2(t)$ (upper curves) and participation number $P(t)$ (lower   curves), (c) the logarithm of the normalized energy distribution
  $E_i$ at time $t_f=10^7$ as a function of lattice site index $i$, and (d)
  the time evolution of the required CPU time $T_C(t)$ in seconds. The
  straight lines in (b) guide the eye for slopes $1/3$ (dashed line)
  and $1/6$ (solid line).  Panels (a), (b) and (d) are in $\log -
  \log$ scale. In panels (a), (b) and (c) the four different curves
  practically overlap each other.  }
\label{fig:1D-4best}
\end{figure}

In Fig.~\ref{fig:1D-4best} we see results obtained for four integrators
of Table \ref{tab:1}, namely $ABA82$ (blue curves), $ABA864$ (red
curves), $SABA_2Y6$ (green curves) and $SABA_2Y8$ (brown curves) for
case B. The values of used $\tau$ for each integrator results in
keeping the absolute relative energy error bounded at the level of
$E_r \approx 10^{-5}$ [Fig.~\ref{fig:1D-4best}(a)]. All integrators
reproduce correctly the dynamical evolution of the system as the
produced results for $m_2(t)$ [upper curves of
  Fig.~\ref{fig:1D-4best}(b)] and $P(t)$ [lower curves of
  Fig.~\ref{fig:1D-4best}(b)], as well as the normalized energy
profiles $E_i$ at $t_f=10^7$ [Fig.~\ref{fig:1D-4best}(c)] practically
overlap. It is worth noting that the results of
Fig.~\ref{fig:1D-4best}(c) show that the second moment and the
participation number of the produced wave packet eventually grow as
$m_2(t) \propto t^{1/3}$ and $P(t) \propto t^{1/6}$ respectively, in
accordance to previously published works
\cite{FKS09,SKKF09,LBKSF10,BLSKF11}. In Fig.~\ref{fig:1D-4best}(d) we
show the required CPU time $T_C$ needed by each SI for this
simulation. From this figure it becomes obvious that the $ABA864$ scheme has
the best performance as it requires the least CPU time.

In Fig.~\ref{fig:1D-4best} we presented results for the best
performing integrators for each order reported in Table
\ref{tab:1}. We note that although the results of Table \ref{tab:1}
and Fig.~\ref{fig:1D-4best} were obtained for the excitation of case
B, the SIs have similar behaviors for all the other studied cases.

\begin{figure}
\centerline{
\includegraphics[scale=1.0]{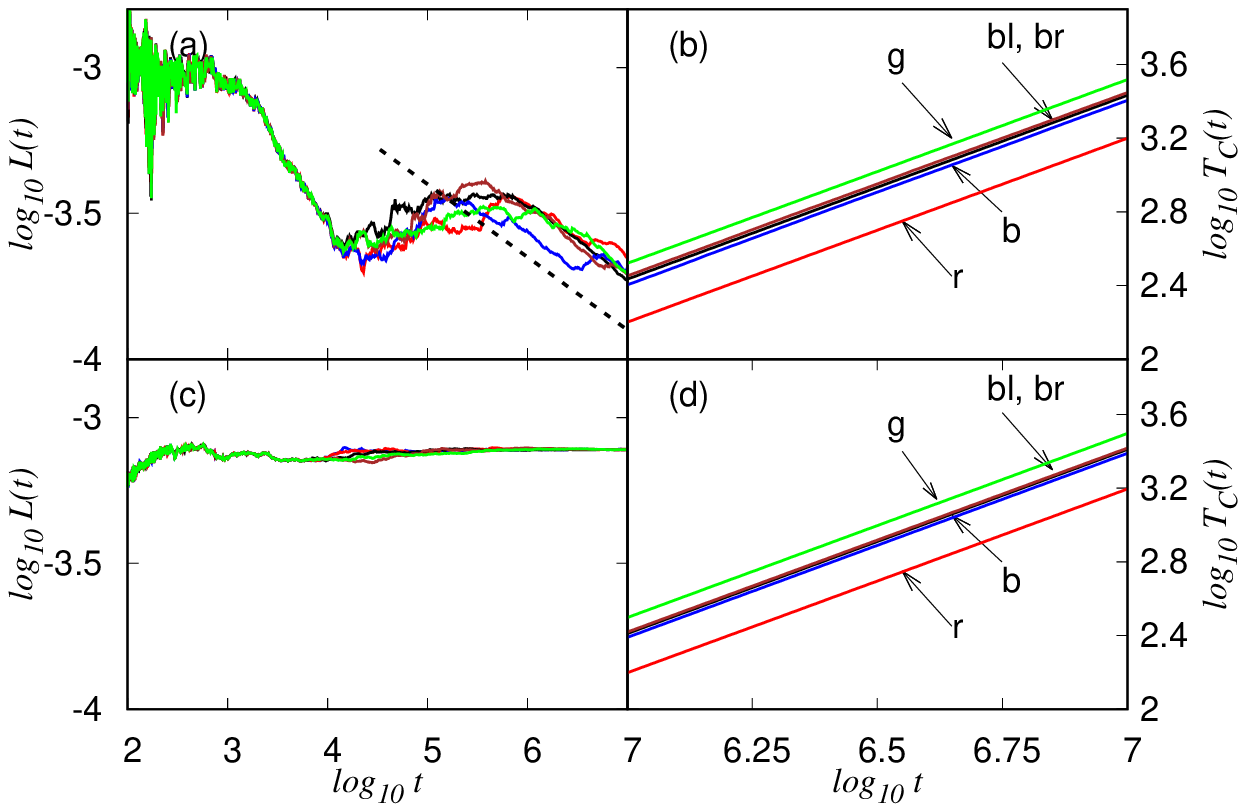}
}
\caption{Results obtained by the integration of the variational
  equations of Hamiltonian (\ref{eq:ham1}) for cases B [panels (a),
    (b)] and E [panels (c), (d)] through the application of the SIs
  $ABA864$, $ABAH864$, $SABA_2Y6$, $s9SABA_26$, $ABA864Y6$  [(r)
    red; (b) blue; (bl) black; (br) brown; (g) green]: the time
  evolution of $L(t)$ (\ref{eq:Lyap}) [(a) and (c)], and of the
  required CPU time $T_C(t)$ [(b) and (d)]. All panels are in $\log -
  \log$ scale. The straight dashed line in (a) guides the eye for
  slope $-1/4$. The curves in (c) practically overlap, as well as the curves for $SABA_2Y6$, $s9SABA_26$ in (b) and (d).}
\label{fig:1D-5best_LEs}
\end{figure}

From the results of Table \ref{tab:1} we see that the SIs exhibiting
the best performance (in descending order of efficiency) are: the
fourth order schemes $ABA864$, $ABAH864$ and the sixth order schemes
$SABA_2Y6$, $s9SABA_26$, $ABA864Y6$. In Fig.~\ref{fig:1D-5best_LEs} we
present results based on the numerical solution of the variational
equations of Hamiltonian (\ref{eq:ham1}) which are obtained by these
five integrators for the weak chaos case B
[Figs.~\ref{fig:1D-5best_LEs}(a) and (b)], as well as the extended
excitation of case E [Figs.~\ref{fig:1D-5best_LEs}(c) and (d)]. From
Figs.~\ref{fig:1D-5best_LEs}(a) and (c) we see that the time evolution
of the finite time mLCE $L(t)$ (\ref{eq:Lyap}) is qualitative the same
for all these SIs. We note that in the weak chaos case B
[Fig.~\ref{fig:1D-5best_LEs}(a)] the $L(t)$ eventually tends to
decrease in a way which is very similar to the law $L(t)\propto
t^{-1/4}$ reported in \cite{SGF13}. This law is different than the
$L(t)\propto t^{-1}$ behavior seen for regular orbits, denoting that
the strength of chaoticity decreases as the wave packet spreads
without showing any sign of crossover to regular dynamics
\cite{SGF13}. On the other hand, for the fully chaotic case E where
all sites are initially excited $L(t)$ shows the typical behavior of
chaos as it saturates quite fast to a constant positive value
[Fig.~\ref{fig:1D-5best_LEs}(c)]. The similarities of Figs.~\ref{fig:1D-5best_LEs}(b) and (d) clearly show that the performance of the integration schemes does not depend on the system's initial conditions and dynamical regime.

In Table \ref{tab:1} only 23 SIs of the 33 considered schemes are
reported because the remaining 10 SIs ($ABA82Y4Y6$, $ABA82Y6$,
$s11SABA_26$, $s11ABA82\_6$ of order six and $SABA_2Y8$, $ABA82Y8$,
$s15SABA_28$, $s19SABA_28$, $s15ABA82\_8$, $s19ABA82\_8$ of order
eight) are unstable. This means that in order to get $E_r \approx
10^{-5}$ they require a rather large integration time step $\tau$,
which is not appropriate for them because they fail to keep the values
of $E_r(t)$ bounded. Requiring a lower bounding value for $E_r$,
e.g.~$E_r \approx 10^{-8}$ (which might not be necessarily needed in
general investigations of disordered lattice dynamics), leads to
smaller values of $\tau$ for which also these integrators keep the
$E_r$ values bounded. We note here that all SIs listed in Table
\ref{tab:1} can achieve $E_r \approx 10^{-8}$ by appropriately
lowering their integration time step.

\begin{figure}
\centerline{
\includegraphics[scale=1.0]{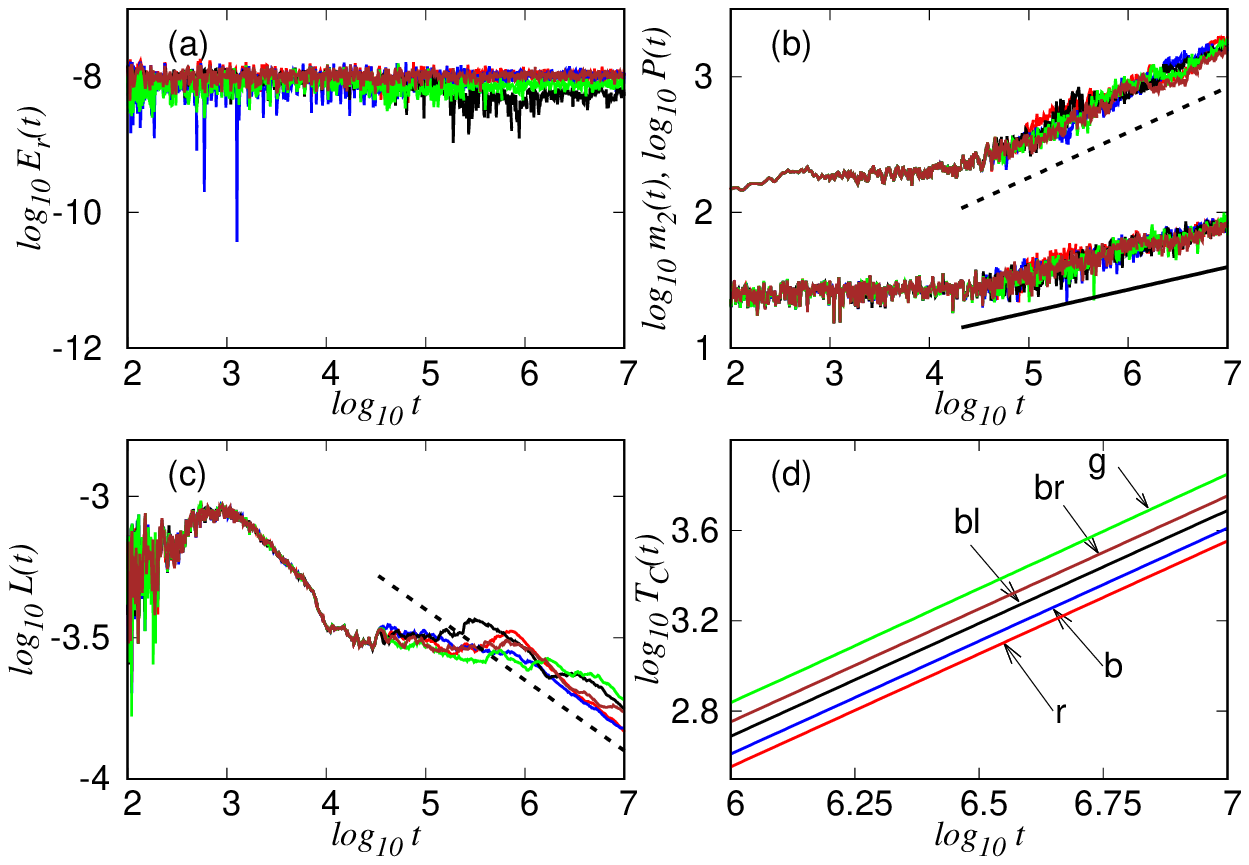}
}
\caption{Results for the integration of case B of Hamiltonian
  (\ref{eq:ham1}) by the SIs $s11ABA82\_6$ (order six) for
  $\tau=0.42$, $s15ABA82\_8$ (order eight) for $\tau=0.48$,
  $s19ABA82\_8$ (order eight) for $\tau=0.59$, $SABA_2Y6$ (order six)
  for $\tau=0.18$ and $s9ABA82\_6$ (order six) for $\tau=0.50$ [(r)
    red; (b) blue; (bl) black; (br) brown; (g) green]: the time
  evolution of the logarithm of (a) $E_r(t)$, (b) $m_2(t)$ (upper
  curves) and $P(t)$ (lower curves), (c) $L(t)$ and (d) $T_C(t)$. All panels are in $\log -
  \log$ scale. The
  straight lines in (b) guide the eye for slopes $1/3$ (dashed line)
  and $1/6$ (solid line), while in (c) the straight dashed line
  corresponds to slope $-1/4$.  }
\label{fig:1D-5best_low}
\end{figure}

By performing a similar analysis to the one presented in
Figs.~\ref{fig:1D-4best} and \ref{fig:1D-5best_LEs} for $E_r \approx
10^{-8}$ we find that the five best performing SIs (in descending
order) are $s11ABA82\_6$ of order six, $s15ABA82\_8$, $s19ABA82\_8$ of
order eight and $SABA_2Y6$, $s9ABA82\_6$ of order six. In
Fig.~\ref{fig:1D-5best_low} we present results obtained by these five
schemes for case B. From Fig.~\ref{fig:1D-5best_low}(a) we see that
the integration time step $\tau$ for each integrator was chosen so
that $E_r \approx 10^{-8}$. All integrators succeeded in capturing
the correct time evolution of $m_2(t)$, $P(t)$
[Fig.~\ref{fig:1D-5best_low}(b)] and $L(t)$
[Fig.~\ref{fig:1D-5best_low}(c)] by producing results similar to the
ones reported if Fig.~\ref{fig:1D-4best}(b) and
Fig.~\ref{fig:1D-5best_LEs}(a) respectively. From the results of
Fig.~\ref{fig:1D-5best_low}(d) we see that all five SIs require more
CPU time than the best five schemes used to obtain $E_r \approx
10^{-5}$ [Fig.~\ref{fig:1D-5best_low}(b)].

\subsection{Two-dimensional KG model}
\label{sec:2d}

We also study the performance of the various SIs for the case of the 2D KG model (\ref{eq:ham2}). For our simulations we consider a two-dimensional $N \times M$ lattice for $N=M=200$ (note that this corresponds to a Hamiltonian system with 40000 degrees of freedom!), create a disorder realization by attributing some random values to the $\epsilon_{i,j}$ parameters in (\ref{eq:ham2}), which remain constant in our numerical experiments, and follow the evolution of energy excitations and  deviation vectors  up to $t_f=10^6$.

By solving the system's equations of motion we keep track of the energy distribution characteristics for three different cases of initial excitations. In particular, we consider the following cases.\\
Case I: we perform a
single site excitation of a site at the center of the lattice for
$W=10$ and total energy $H_{2I}=0.3$.\\
Case II: similar to case I but for $W=10$ and total energy $H_{2II}=2.0$.\\
Case III: all sites are
initially excited, having the same amount of initial energy, for $W=10$ and total energy $H_{2III}=10$.\\
We note that case I corresponds to the system's weak chaos regime and case II to the selftrapping regime, while case III represents a general, extended initial excitation. The dynamics of cases I and II was studied in \cite{LBF12}.

As in the 1D case we consider two-dimensional normalized energy distributions
\begin{equation}
\label{eq:Eij}
\begin{array}{lll}
\displaystyle E_{i,j} & = & \Bigg\{ \displaystyle \frac{p^2_{i,j}}{2} + \frac{\epsilon_{i,j} q^2_{i,j} }{2}  +
\frac{   q_{i,j}  ^{4}}{4} + \frac{1}{4W} \bigg[ (q_{i+1,j} -q_{i,j})^2 + (q_{i,j+1} - q_{i,j})^2  \\
      & & \displaystyle  +(q_{i,j} - q_{i-1,j})^2 + (q_{i,j} - q_{i,j-1})^2 \bigg] \Bigg\} / H_2\, , \,\,\, i=1,\ldots,N, \,\, j=1,\ldots,M,
\end{array}
\end{equation}
and evaluate their second moment $m_2=\sum_{i}^{N} \sum_{j}^{M} \left\| (i,j)-(\bar{i},\bar{j}) \right\|^2 E_{i,j}$ and participation number $P=\left( \sum_{i}^{N} \sum_{j}^{M} E_{i,j} ^2\right)^{-1}$,  with $(\bar{i},\bar{j}) = \sum_{i=1}^N \sum_{j=1}^M (i,j) E_{i,j}$. In addition, by solving the system's variational equations we follow the time evolution of deviation vectors and compute, to the best of our knowledge for the first time, the finite time mLCE (\ref{eq:Lyap}) for this model. The initial deviation vector used has random, nonzero coordinates only at a square of $4 \times 4 = 16$ sites at the center of the lattice. Due to the complexity of the set of equations of motion and the variational ones we explicitly present in the Appendix \ref{sec:ap} the form of operators $e^{\tau L_{AV_2}}$ (\ref{eq:AV2}) and $e^{\tau L_{BV_2}}$ (\ref{eq:BV2}) used in the various symplectic integration schemes, hoping that they will be useful for researchers working on the dynamics of 2D KG models or similar systems.

\begin{figure}
\centerline{
\includegraphics[scale=1.0]{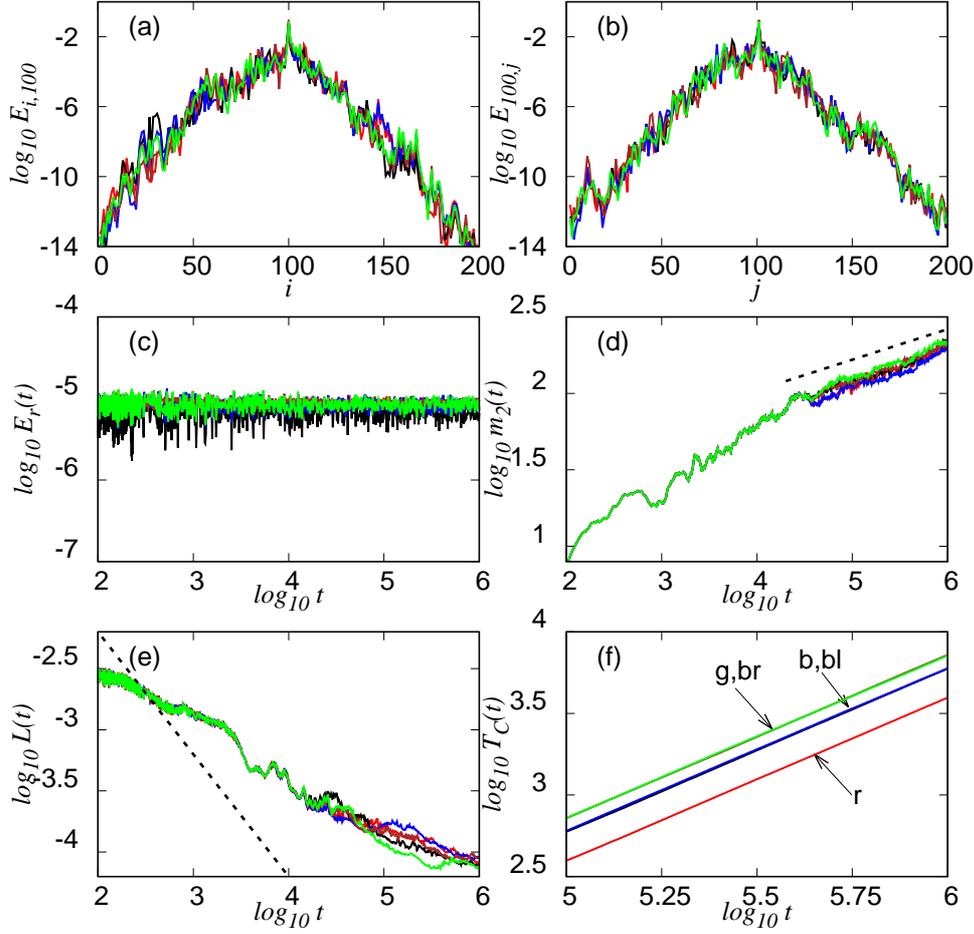}
}
\caption{Results for the integration of case I of Hamiltonian
  (\ref{eq:ham2}) by the SIs $ABA864$ (order four) for $\tau=0.56$, $ABAH864$ (order four) for $\tau=0.40$, $SABA_2Y6$ (order six) for $\tau=0.60$, $s9SABA_26$ (order six) for $\tau=0.60$, $ABA864Y6$ (order six) for $\tau=0.68$ [(r)
    red; (b) blue; (bl) black; (br) brown; (g) green].  The logarithm of the normalized energy distribution (a)
  $E_{i,100}$  and (b) $E_{100,j}$ at time $t_f=10^6$ as a function of the lattice site index $i$ and $j$ respectively.
The time
  evolution of the logarithm of (c) $E_r(t)$, (d) $m_2(t)$, (e) $L(t)$ and (f) $T_C(t)$. Panels (c), (d), (e) and (f)
  are in $\log -
  \log$ scale. The
  straight dashed line in (d) guides the eye for slope $0.2$, while in (e) for slope $-1$. In (f) the curves for $ABAH864$, $SABA_2Y6$ and $s9SABA_26$, $ABA864Y6$ practically overlap each other.}
\label{fig:2D}
\end{figure}

In Fig.~\ref{fig:2D} we present results obtained for case I by implementing the five best performing SIs among the studied schemes, when an absolute relative error $E_r\approx 10^{-5}$ was required [Fig.~\ref{fig:2D}(c)]. These are the same five SIs which exhibited the best numerical performance also for the 1D KG system (Fig.~\ref{fig:1D-5best_LEs}): $ABA864$, $ABAH864$  of order four, and $SABA_2Y6$, $s9SABA_26$, $ABA864Y6$ of order six. All these integrators succeeded in correctly capturing the dynamics of the system. 

As the energy propagation takes place on a two-dimensional plane it is difficult to visualize in a comparative way the energy distributions obtained by the different integrators.  For this reason we present in
Fig.~\ref{fig:2D}(a) the energy profiles along the $i$ axis for sites having $j=100$ and in  Fig.~\ref{fig:2D}(b) along the $j$ axis for sites with $i=100$ at $t_f=10^6$. We clearly see that the profiles produced by the different integrators practically coincide. Another piece of evidence that all schemes provide the same results is the fact that very nearly they produce the same time evolution of $m_2(t)$ [Fig.~\ref{fig:2D}(d)], $P(t)$ (not shown) and $L(t)$ [Fig.~\ref{fig:2D}(e)]. From  Fig.~\ref{fig:2D}(d) we see that eventually $m_2(t) \propto t^{0.2}$ in agreement with the results presented in \cite{LBF12} for the weak chaos case.

The evolution of $L(t)$ in Fig.~\ref{fig:2D}(e) shows a power law decrease  but with a rate which is completely different than the $L(t) \propto t^{-1}$ decay [dashed line in Fig.~\ref{fig:2D}(e)] observed in the case of regular motion. This behavior indicates that the motion becomes less chaotic in time but without showing any sign of a crossover to regular dynamics. This is similar to what has been observed in the case of weak chaos of the 1D KG system \cite{SGF13} [see also Figs.~\ref{fig:1D-5best_LEs}(a) and \ref{fig:1D-5best_low}(c)], where $L(t) \propto t^{-1/4}$. Fig.~\ref{fig:2D}(e) suggests that $L(t) \propto t^{-\nu}$ with $\nu> 0$ and $\nu \neq 1$, also for the 2D KG case, but the presented results are not enough to provide an accurate estimation for $\nu$ as they are based only on one disorder realization. A more detailed investigation of the evolution of $L(t)$ for larger integration times and many more disorder realizations and values of the system's parameters (e.g.~total energy and disorder strength) is needed. We plan to address this issue in a future publication.

From Fig.~\ref{fig:2D}(f) we see that, as in the case of the 1D KG model (Fig.~\ref{fig:1D-5best_LEs}), the best performing scheme is the SI $ABA864$ followed by the $ABAH864$, $SABA_2Y6$ and $s9SABA_26$, $ABA864Y6$  schemes, which behave similarly as their $T_C(t)$ vs.~time curves practically overlap. We also note that similar results to the ones presented in Fig.~\ref{fig:2D} were also obtained for cases II and III, showing again that the most efficient integration scheme is the fourth order SI $ABA864$.

\section{Summary and conclusions}
\label{sec:conclusions}

We carried out a detailed analysis of the performance of several SIs, with orders ranging from two up to eight,  for the long time integration of the equations of motion and the variational equations of the 1D (\ref{eq:ham1}) and the 2D (\ref{eq:ham2}) KG models.
By performing extensive numerical simulations we investigated the ability of 33 SIs to correctly capture the characteristics of energy propagation induced by different initial excitations in these models,
as well as to accurately quantify the systems' chaoticity. In particular, we followed the time evolution
of energy distributions and computed their second moment and participation number along with the corresponding finite time mLCE,  by implementing each one of these SIs, registering also the CPU time they required. Our results show that the behavior of the tested SIs does not depend on the nature of the used initial excitations or  the considered dynamical regime of the two lattice models.

For both models the integrators  $ABA864$, $ABAH864$ of
order four and $SABA_2Y6$, $s9SABA_26$, $ABA864Y6$ of order six, exhibited the best performance when a moderate accuracy in the
numerical conservation of the system's energy was required (i.e.~the corresponding absolute relative energy error was $E_r\approx 10^{-5}$), with $ABA864$ \cite{BCFLMM13} being always the most efficient one as it  required the least CPU time. This is indeed a very efficient SI as also three part split symplectic schemes based on it showed the best performance among several SIs tested for the integration of the DNLS system \cite{SGBPE14,GMS16}. Thus, we propose that the $ABA864$ SI should be preferred for the long time integration of the 1D and 2D KG models over SIs of the $SABA$ family \cite{LR01} which have been extensively used to date for such studies \cite{FKS09,SKKF09,SF10,LBKSF10,BLSKF11,LBF12}.
This is a basic outcome of our study, which can be of practical importance for researchers working on lattice dynamics of disordered and non-disordered systems.

Many of the considered sixth and eight order SIs were not able to produce reliable results for  $E_r\approx 10^{-5}$
because the required integration time step $\tau$ needed for that purpose was rather high and made them unstable. These
higher order SIs can be used in cases where an even better accuracy is required than the typically acceptable level of  $E_r\approx 10^{-5}$ used in disordered lattice studies. For $E_r\approx 10^{-8}$ the best performing schemes were the SIs
$s11ABA82\_6$, $SABA_2Y6$, $s9ABA82\_6$ of order six and $s15ABA82\_8$, $s19ABA82\_8$ of
order eight, with $s11ABA82\_6$ being the most efficient one.

In our study we paid much attention to the integration of the variational equations by SIs based on the `tangent map method' \cite{SG10,GS11,GES12}, because through their solution we can evaluate the mLCE and quantify the system's chaoticity. For this reason we provide in the Appendix the explicit formulas of the operators needed for this task, both for the rather simple 1D KG system and the more complicated case of the 2D KG model. Our results indicate that it is possible to obtain the long time evolution of the finite time mLCE in feasible CPU times for both the 1D and the 2D KG systems. For the 2D KG system in particular, we evaluated the mLCE  obtaining some numerical evidences (to the best of our knowledge for the first time) that in the weak chaos regime the mLCE  decreases to zero following a power law which is different than the $t^{-1}$ law encountered for regular orbits. This behavior is similar to what has been observed also in the 1D KG case \cite{SGF13}.  A more detailed investigation of the behavior of the mLCE
for the different dynamical regimes appearing in the 2D KG model is needed in order to derive reliable conclusions for the system's chaoticity, something we plan to address in the near future. The SIs presented in our work can  facilitate the realization of this goal as they managed to speed up considerably the needed computations.

\begin{acknowledgement}
Ch.~S.~was supported by the National Research Foundation of South
Africa (Incentive Funding for Rated Researchers, IFRR and Competitive
Programme for Rated Researchers, CPRR). We thank J.~D.~Bodyfelt for
fruitful discussions and K.~B.~Mfumadi for checking some of our
computations.

\end{acknowledgement}

\appendix
\section{The $e^{\tau L_{AV}}$ and $e^{\tau L_{BV}}$ operators}
\label{sec:ap}

We present here the explicit form of operators $e^{\tau L_{AV}}$ and
$e^{\tau L_{BV}}$ used for the time propagation of an orbit and a
deviation vector with initial conditions $(\vec{q}, \vec{p},
\vec{\delta q}, \vec{\delta p})$ at time $t_0$ to their final values
$(\vec{q}', \vec{p}', \vec{\delta q}', \vec{\delta p}')$ at time $t_0+
\tau$ for Hamiltonians (\ref{eq:ham1}) and (\ref{eq:ham2}).

\subsection{The one-dimensional KG model}
\label{sec:ap1}

The equations of motion of the 1D KG model (\ref{eq:ham1}) are
\begin{equation}
\displaystyle
\begin{array}{lll}
      \displaystyle \frac{dq_i}{dt}&=& \displaystyle p_{i},
      \,\,\,\mbox{for}\,\,\, 1\leq i\leq N\\ \displaystyle
      \frac{dp_1}{dt}&=& \displaystyle - \left[\epsilon_{1}q_{1} +
        q_{1}^3 +
        \frac{1}{W}\left(2q_{1}-q_{2}\right)\right]\\ \displaystyle
      \frac{dp_i}{dt}&=& \displaystyle - \left[\epsilon_{i}q_{i} +
        q_{i}^3 + \frac{1}{W}\left(2q_{i}-q_{i-1}
        -q_{i+1}\right)\right], \,\,\,\mbox{for}\,\,\, 2\leq i\leq
      N-1\\ \displaystyle \frac{dp_N}{dt}&=& \displaystyle -
      \left[\epsilon_{N}q_{N} + q_{N}^3 +
        \frac{1}{W}\left(2q_{N}-q_{N-1}\right)\right],
\end{array}
\label{eq:1eqmot}
\end{equation}
while the corresponding variational equations (\ref{eq:var}) have the form
\begin{equation}
\displaystyle
\begin{array}{lll}
      \displaystyle \frac{d\delta q_i}{dt}&=& \displaystyle \delta p_{i}, \,\,\,\mbox{for}\,\,\,
      1\leq i\leq N \\
      \displaystyle \frac{d\delta p_1}{dt}&=& \displaystyle - \left[
      \delta q_{1}\left(\epsilon_{1}+3q_{1}^2\right)
      + \frac{1}{W}\left(2\delta q_{1}-\delta q_{2}\right)\right]\\
      \displaystyle \frac{d\delta p_i}{dt}&=& \displaystyle - \left[
      \delta q_{i}\left(\epsilon_{i}+3q_{i}^2\right)
      + \frac{1}{W}\left(2\delta q_{i}-\delta q_{i-1}
      -\delta q_{i+1}\right)\right], \,\,\,\mbox{for}\,\,\,  2\leq i\leq N-1\\
      \displaystyle \frac{d\delta p_N}{dt}&=& \displaystyle - \left[
      \delta q_{N}\left(\epsilon_{N}+3q_{N}^2\right)
      + \frac{1}{W}\left(2\delta q_{N}-\delta q_{N-1}\right)\right].
\end{array}
\label{eq:1vareq}
\end{equation}

In order to implement the SIs of Sect.~\ref{sec:si} for the
simultaneous integration of equations (\ref{eq:1eqmot}) and
(\ref{eq:1vareq}) we split Hamiltonian (\ref{eq:ham1}) in two
integrable parts
\begin{equation}
\label{eq:ham1_split}
 A_1(\vec{p}) = \sum_{i=1}^N \frac{p_i^2}{2}\, , \,\,\,\,\,\,
   B_1(\vec{q}) =\sum_{i=1}^N \left[
   \frac{\epsilon_i}{2}q_i^2 + \frac{q_i^4 }{4}+
   \frac{1}{2W}\left(q_{i+1} - q_i\right)^2 \right],
\end{equation}
i.e.~the system's kinetic and potential energy respectively. The
solution of the Hamilton equations of motion and the variational ones
for Hamiltonians $A_1$ and $B_1$ are obtained through the action of
the operators
\begin{equation}
e^{\tau L_{AV_1}}: \left\{
\begin{array}{lll}
      \displaystyle q'_{i}&=& \displaystyle q_{i} + \tau p_{i} \\
      \displaystyle p'_{i}&=& \displaystyle p_{i}\\
      \displaystyle \delta q'_{i}&=& \displaystyle \delta q_{i} + \tau \delta p_{i} \\
      \displaystyle \delta p'_{i}&=& \displaystyle \delta p_{i}
\end{array}
\right. , \,\,\,\mbox{for}\,\,\,  1\leq i\leq N,
\label{eq:AV1}
\end{equation}
and
\begin{equation}
e^{\tau L_{BV_1}}: \left\{
\begin{array}{lll}
      \displaystyle q'_{i}&=& \displaystyle q_{i}, \,\,\,\mbox{for}\,\,\,  1\leq i\leq N\\
      \displaystyle p'_{1}&=& \displaystyle p_{1} - \tau\left[\epsilon_{1}q_{1} +
      q_{1}^3 + \frac{1}{W}\left(2q_{1}-q_{2}\right)\right]\\
      \displaystyle p'_{i}&=& \displaystyle p_{i} - \tau\left[\epsilon_{i}q_{i} +
      q_{i}^3 + \frac{1}{W}\left(2q_{i}-q_{i-1}
      -q_{i+1}\right)\right], \,\,\,\mbox{for}\,\,\,  2\leq i\leq N-1\\
      \displaystyle p'_{N}&=& \displaystyle p_{N} - \tau\left[\epsilon_{N}q_{N} +
      q_{N}^3 + \frac{1}{W}\left(2q_{N}-q_{N-1}\right)\right]\\
      & & \\
      \displaystyle \delta q'_{i}&=& \displaystyle \delta q_{i}, \,\,\,\mbox{for}\,\,\,  1\leq i\leq N\\
      \displaystyle \delta p'_{1}&=& \displaystyle \delta p_{1} - \tau\left[
      \delta q_{1}\left(\epsilon_{1}+3q_{1}^2\right)
      + \frac{1}{W}\left(2\delta q_{1}-\delta q_{2}\right)\right]\\
      \displaystyle \delta p'_{i}&=& \displaystyle \delta p_{i} - \tau\left[
      \delta q_{i}\left(\epsilon_{i}+3q_{i}^2\right)
      + \frac{1}{W}\left(2\delta q_{i}-\delta q_{i-1}
      -\delta q_{i+1}\right)\right], \,\,\,\mbox{for}\,\,\,  2\leq i\leq N-1\\
      \displaystyle \delta p'_{N}&=& \displaystyle \delta p_{N} - \tau\left[
      \delta q_{N}\left(\epsilon_{N}+3q_{N}^2\right)
      + \frac{1}{W}\left(2\delta q_{N}-\delta q_{N-1}\right)\right].
\end{array}
\right.
\label{eq:BV1}
\end{equation}
Note that the first half of the equations of operators (\ref{eq:AV1})
and (\ref{eq:BV1}) correspond respectively to the operators $e^{\tau L_{A_1}}$ and
$e^{\tau L_{B_1}}$  needed for the integration of only the
system's equations of motion.

\subsection{The two-dimensional KG model}
\label{sec:ap2}

The 2D KG Hamiltonian (\ref{eq:ham2}) can also be written as the sum
of two integrable systems: the kinetic energy $A_2(\vec{p})$ and the
potential energy $B_2(\vec{q})$. In this case the propagation
operators for the solution of the equations of motion and the
variational equations have more complicated forms with respect to
the 1D case and are given by the following expressions
\begin{equation}
e^{\tau L_{AV_2}}: \left\{
\begin{array}{lll}
      \displaystyle q'_{i,j}&=& \displaystyle q_{i,j} + \tau p_{i,j} \\
      \displaystyle p'_{i,j}&=& \displaystyle p_{i,j}\\
      \displaystyle \delta q'_{i,j}&=& \displaystyle \delta q_{i,j} + \tau \delta p_{i,j} \\
      \delta p'_{i,j}&=& \displaystyle \delta p_{i,j}
\end{array}
\right. , \,\,\,\mbox{for}\,\,\,  1\leq i\leq N, \,\, 1\leq j\leq M,
\label{eq:AV2}
\end{equation}
and
\begin{equation}
e^{\tau L_{BV_2}}: \left\{
\begin{array}{lll}
      \displaystyle q'_{i,j}&=& \displaystyle q_{i,j}, \,\,\,\mbox{for}\,\,\,  1\leq i\leq N, \,\, 1\leq j\leq M\\
      \displaystyle p'_{1,1}&=& \displaystyle p_{1,1} - \tau\left[\epsilon_{1,1}q_{1,1} + q_{1,1}^3 + \frac{1}{W}\left(4q_{1,1}-q_{2,1}-q_{1,2}\right)\right]\\
      \displaystyle p'_{1,M}&=& \displaystyle p_{1,M} - \tau\left[\epsilon_{1,M}q_{1,M} +  q_{1,M}^3 + \frac{1}{W}\left(4q_{1,M}-q_{1,M-1}-q_{2,M}\right)\right]\\
      \displaystyle p'_{N,1}&=& \displaystyle p_{N,1} - \tau\left[\epsilon_{N,1}q_{N,1} +  q_{N,1}^3 + \frac{1}{W}\left(4q_{N,1}-q_{N-1,1}
      -q_{N,2}\right)\right]\\
      \displaystyle p'_{N,M}&=& \displaystyle p_{N,M} - \tau\left[\epsilon_{N,M}q_{N,M} +  q_{N,M}^3 + \frac{1}{W}\left(4q_{N,M}-q_{N-1,M}
      -q_{N,M-1}\right)\right]\\
      \displaystyle p'_{i,j}&=& \displaystyle p_{i,j} - \tau\left[\epsilon_{i,j}q_{i,j} +  q_{i,j}^3 +\frac{1}{W}\left(4q_{i,j}- q_{i-1,j}  -q_{i,j-1}\right. \right.\\
       & & \displaystyle  \left. \left. -q_{i+1,j}-q_{i,j+1}\right) \bigg] \right. , \,\,\,\mbox{for}\,\,\,  2\leq i\leq N-1,\,\,2\leq j\leq M-1\\
       \displaystyle p'_{i,1}&=& \displaystyle p_{i,1} - \tau\left[\epsilon_{i,1}q_{i,1} +  q_{i,1}^3 + \frac{1}{W}\left(4q_{i,1}-q_{i-1,1} -q_{i+1,1}-q_{i,2}\right)\right], \,\,\,\mbox{for}\,\,\,   2\leq i\leq N-1\\
      \displaystyle p'_{i,M}&=& \displaystyle p_{i,M} - \tau\left[\epsilon_{i,M}q_{i,M} +  q_{i,M}^3 + \frac{1}{W}\left(4q_{i,M}-q_{i-1,M}  -q_{i,M-1}-q_{i+1,M}\right)\right], \,\,\,\mbox{for}\,\,\,   2\leq i\leq N-1\\
      \displaystyle p'_{1,j}&=& \displaystyle p_{1,j} - \tau\left[\epsilon_{1,j}q_{1,j} +  q_{1,j}^3 +\frac{1}{W}\left(4q_{1,j}-q_{1,j-1}-  q_{2,j}-q_{1,j+1}\right)\right], \,\,\,\mbox{for}\,\,\,  2\leq j\leq M-1\\
      \displaystyle p'_{N,j}&=& \displaystyle p_{N,j} - \tau\left[\epsilon_{N,j}q_{N,j} +  q_{N,j}^3 +\frac{1}{W}\left(4q_{N,j}-q_{N-1,j}  -q_{N,j-1}-q_{N,j+1}\right)\right], \,\,\,\mbox{for}\,\,\,    2\leq j\leq M-1\\

      & & \\

      \displaystyle \delta q'_{i,j}&=& \displaystyle \delta q_{i,j}, \,\,\,\mbox{for}\,\,\,    1\leq i\leq N,\,\, 1\leq j\leq M\\
      \displaystyle \delta p'_{1,1}&=& \displaystyle \delta p_{1,1} - \tau\left[      \delta q_{1,1}\left(\epsilon_{1,1}+3q_{1,1}^2\right)  +  \frac{1}{W}\left(4\delta q_{1,1}-\delta q_{2,1}- \delta q_{1,2}\right)\right]\\
      \displaystyle \delta p'_{1,M}&=& \displaystyle \delta p_{1,M} - \tau\left[  \delta q_{1,M}\left(\epsilon_{1,M}+3q_{1,M}^2\right)   +  \frac{1}{W}\left(4\delta q_{1,M}-\delta q_{1,M-1}-  \delta q_{2,M}\right)\right]\\
      \displaystyle \delta p'_{N,1}&=& \displaystyle \delta p_{N,1} - \tau\left[  \delta q_{N,1}\left(\epsilon_{N,1}+3q_{N,1}^2\right)    + \frac{1}{W}\left(4\delta q_{N,1}-\delta q_{N-1,1}  -\delta q_{N,2}\right)\right]\\
      \displaystyle \delta p'_{N,M}&=& \displaystyle \delta p_{N,M} - \tau\left[   \delta q_{N,M}\left(\epsilon_{N,M}+3q_{N,M}^2\right)    + \frac{1}{W}\left(4\delta q_{N,M}-\delta q_{N-1,M}   -\delta q_{N,M-1}\right)\right]\\
      \displaystyle \delta p'_{i,j}&=& \displaystyle \delta p_{i,j} - \tau\left[  \delta q_{i,j}\left(\epsilon_{i,j}+3q_{i,j}^2\right)   +\frac{1}{W}\left(4\delta q_{i,j}-\delta q_{i-1,j}  -\delta q_{i,j-1}\right. \right.\\
      & & \displaystyle \left. \left.-\delta q_{i+1,j}-\delta q_{i,j+1}\right) \bigg] \right., \,\,\,\mbox{for}\,\,\,   2\leq i\leq N-1, \,\, 2\leq j\leq M-1 \\
      \displaystyle \delta p'_{i,1}&=& \displaystyle \delta p_{i,1} - \tau\left[   \delta q_{i,1}\left(\epsilon_{i,1}+3q_{i,1}^2\right)    + \frac{1}{W}\left(4\delta q_{i,1}-\delta q_{i-1,1}  -\delta q_{i+1,1}-\delta q_{i,2}\right)\right], \,\,\,\mbox{for}\,\,\,  2\leq i\leq N-1\\
      \displaystyle \delta p'_{i,M}&=& \displaystyle \delta p_{i,M} - \tau\left[   \delta q_{i,M}\left(\epsilon_{i,M}+3q_{i,M}^2\right)   +\frac{1}{W}\left(4\delta q_{i,M}-\delta q_{i-1,M} -\delta q_{i,M-1}-\delta q_{i+1,M}\right)\right], \,\,\,\mbox{for}\,\,\,    2\leq i\leq N-1\\
      \displaystyle \delta p'_{1,j}&=& \displaystyle \delta p_{1,j} - \tau\left[  \delta q_{1,j}\left(\epsilon_{1,j}+3q_{1,j}^2\right)     +\frac{1}{W}\left(4\delta q_{1,j}   -\delta q_{1,j-1}-\delta q_{2,j}-\delta q_{1,j+1}\right)\right], \,\,\,\mbox{for}\,\,\,    2\leq j\leq M-1\\
      \displaystyle \delta p'_{N,j}&=& \displaystyle \delta p_{N,j} - \tau\left[   \delta q_{N,j}\left(\epsilon_{N,j}+3q_{N,j}^2\right)    +\frac{1}{W}\left(4\delta q_{N,j}-\delta q_{N-1,j}  -\delta q_{N,j-1}-\delta q_{N,j+1}\right)\right], \,\,\,\mbox{for}\,\,\,    2\leq j\leq M-1.\\
\end{array}
\right.
\label{eq:BV2}
\end{equation}


\end{document}